\documentclass[twocolumn,showpacs,preprintnumbers,amsmath,amssymb]{revtex4}
\usepackage{graphicx}% Include figure files
\usepackage{dcolumn}% Align table columns on decimal point
\usepackage{bm}% bold math
\usepackage{hyperref}% makes the references hyperlinked
\newcommand{\hodge}[1]{\,*#1}
\newcommand{\lie}{{\cal L}}
\begin{document}
%\preprint{}
\title{de Sitter Thermodynamics: A glimpse into non equilibrium}
\author{Rodrigo Aros}
\affiliation{Departamento de Ciencias F\'{\i}sicas, Universidad Andr\'es Bello,
Av. Republica 252,
Santiago,Chile}

\date{\today}
\begin{abstract}
In this article is shown that the thermodynamical evolution of a Schwarzschild de Sitter space is
the evaporation of its black hole. The result is extended in higher dimensions to Lovelock
theories of gravity with a single positive cosmological constant.
\end{abstract}
\maketitle

\section{Introduction}

The study of the thermodynamics of black holes has been so far the
only windows at hand into the
realms of quantum gravity. One  of the most remarkable results obtained was to
prove that the boundary conditions define the ensemble \cite{Brown:1993bq} in which the black hole
is described. Unfortunately, there are not generical boundary conditions that one can identify with
any particular ensemble, and this must be done case by case. Furthermore, there can be more than a
single set of boundary conditions that yield any ensemble. For instance in \cite{Aros:2005by} was
shown that the boundary conditions which define the canonical ensemble with null and negative
cosmological constants are not related at all.

In \cite{Aros:2005by} the case of positive cosmological constant was excluded and the present
article aims to amend that in part. The simplest case with positive cosmological constant is the
de Sitter space which is the maximally symmetric manifold with positive curvature. It has a
horizon which is an observer depending feature. This is associated with the fact its Euclidean
version is a sphere. Moreover its finite volume has led to conjecture that it could have a finite
number of quantum states \cite{Witten:2001kn}.

However, since maximally symmetric spaces usually are stable, if not plainly ground states, the
presence of a horizon in de Sitter space may be considered in conflict with the idea that a
horizon emits Hawking radiation in an underlaying decaying process. The same discussion for
negative cosmological constants is solved by the presence, or not, of Killing spinors because of
their connection with the definition of a BPS state \cite{Aros:2002rk}. For a positive
cosmological constant such a connection can not be established, although Killing spinors indeed
exist, because the de Sitter group does not have a supersymmetric extension (see for instance
\cite{Freund:1986ws}). Therefore, in order to understand the role of the de Sitter as a ground
state can be useful to study the thermodynamics of black holes with positive cosmological
constant. Even studying the simplest case, the Schwarzschild-dS solution, gives a lot of useful
information. There are several other geometries with positive cosmological constant whose
thermodynamics can be relevant, for a discussion see \cite{Myung:2006tg}, unfortunately most of
them present naked singularities.

From the start the thermodynamics of black holes with positive cosmological constant presents some
novelties. One usually deals with a \textit{single horizon} where somehow to fix a \textit{single
temperature}. For a positive cosmological constant, in even for Schwarzschild-dS solution, the
space where the observers \textit{inhabit} is located between two horizons, and at both a
temperature can be defined \cite{Gomberoff:2003ea,Choudhury:2004ph}. 

In principle the presence of those two horizons with their own temperature defines a
non-equilibrium system, which should evolve. One very interesting
feature of the Schwarzschild-dS black hole geometry is that its black hole horizon can be understood
as made of degrees of freedom borrowed from the cosmological horizon \cite{Astefanesei:2003gw}.
This is in complete agrement with the fact even if one adds cosmological horizon and
the black hole horizon entropies still the result is smaller than the entropy of de Sitter space,
defined as usual as proportional to the respective areas. Remarkably, with only this in mind one can
predict, using the usual rule that systems evolve into larger entropy configurations, that
Schwarzschild-dS space should evolve into de Sitter space. In this paper that evolution is discussed
on some general grounds. The final result of this article is that the quasi statical thermodynamical
evolution determines the complete evaporation of the black hole of the Schwarzschild-dS solution,
leaving behind, in principle, a de Sitter space.

In this article the positive cosmological constant will be considered fixed, though it is known
that even the cosmological constant can evolve \cite{Gomberoff:2003zh}.

\subsection*{Thermodynamics reviewed}
Before to proceed to the next sections is worth to recall some notions of black hole
thermodynamics.

Thermodynamics has two fundamental laws which are satisfied by every known physical system,
therefore one should expect that they be satisfied by black holes as well. Above all stands the
conservation of energy, known as the first law of thermodynamics,
\begin{equation}\label{FirstLaw}
    dE = dQ + \sum_{i} \mu_{i} dJ^{i},
\end{equation}
where $dQ$ stands for the differential of heat, $J^{i}$ are some extensive charges, as angular
momenta, and $\mu_{i}$ their associated extensive potentials (For gravity see for instance
\cite{Hawking:1998kw}). This law can be even used to recover the gravitational equations for black
holes, see for instance \cite{Paranjape:2006ca}.

The other fundamental law is the so called second law of thermodynamics
\begin{equation}\label{SecondLaw}
    \sum_{a} dS_{a} \geq 0,
\end{equation}
which states that in the evolution of a composed system the total change of entropy is always
positive or null.

The suitability of the other two laws of thermodynamics in black hole physics is not so clear. The
zero law, which states that two systems in contact must reach thermal
equilibrium, needs at least that the heat capacities be positive. This can fail in gravity (for
instance it fails in Newtonian gravity). Finally, the third law of thermodynamics also
represents an open
question, since to step from a non vanishing into vanishing temperature black hole is not a smooth
geometrical process and represents a change of topology (for a discussion see
\cite{Belgiorno:2002pm}).

\section{A bounded space}

For simplicity the discussion will be centered on the Schwarzschild-dS solution. The extension to
Kerr-dS is discussed in appendix \ref{KerrSection}. The Schwarzschild-dS line element in $d$
dimensions reads
\begin{equation}\label{SchwdeSitter}
ds^{2} = -f(r)^2 dt^{2} + \frac{1}{f(r)^2}dr^{2} + r^{2} d\Omega^{2}_{d-2}
\end{equation}
where
\begin{equation}\label{fr}
f(r)^2=1 -\frac{r^2}{l^{2}} - \frac{2 M G_{1}}{r^{d-3}}
\end{equation}
and $d\Omega^{2}_{d-2}$ the line element of $S^{d-2}$.

For $d>3$ one can notice that $f(r)^ 2$ may have none, one, two positive roots depending on the
value of $M$. To avoid naked singularities $M \geq 0$. This article deals with the case with two
positive roots and the evolution of space defined between those two radii. In that case the
largest root, called $r_{++}$, defines the radius of a cosmological horizon. The smallest root,
called $r_{+}$, defines the radius of a black hole horizon.

Since in Eq.(\ref{fr}) $M$ is the only free parameter, $r_{+}$ and $r_{++}$ can not be mutually
independent. For instance, in four dimensions they satisfy
\[
(r_{+}+r_{++})^2 - r_{+}r_{++} = l^2.
\]
In higher dimension there are analogous relations (see Eq.(\ref{dDimensionalrelation})). The ranges
of those radii are given by $0 \leq r_{+} < r_{0}$ and $r_{0}< r_{++} \leq l$ where
\[
r_{0}= l \sqrt{\frac{d-3}{d-1}}.
\]
Finally $M$ is restricted by $0 \leq M < M_{max}$ with
\[
G_{1}M_{max} = \frac{l^{d-3}}{d-3}\left(\frac{d-3}{d-1}\right)^{\frac{d-1}{2}}.
\]

One can notice that, although the Schwarzschild de Sitter solution shares some of the basic
structures of de Sitter space, its cosmological horizon is not observer dependent, but a
\textit{real} horizon. This is due to
the presence of the black hole, which breaks the global symmetries involving the \textit{radial
direction}. In fact the geometry is given by $\mathcal{M}=\mathbb{R}\times \Sigma$ where $\Sigma$
is a $d-1$-dimensional spacelike hypersurface and $\mathbb{R}$ stands for the time direction. The
boundary is therefore given by  $\partial \mathcal{M}=\mathbb{R}\times \partial \Sigma$, where
$\partial\Sigma=\partial \Sigma_{+} \oplus \partial \Sigma_{++}$. $\partial \Sigma_{+}$ and
$\partial \Sigma_{++}$ stands for the black hole and cosmological horizons respectively. These
definitions can be extended to Kerr-dS in a natural way. See appendix \ref{KerrSection}.

In three dimensions the structure is different. In this case the solution is
locally a de Sitter space and there is a single cosmological horizon. For any value of $M \neq 0$
the space presents a conical singulary at $r=0$ which increases with $M$. For $G_{1}M \geq 1/2$ the
solution (\ref{fr}) becomes ill defined and $M<0$ is forbidden since it introduces conical
singularities with an excess of angle. Therefore $0\leq G_{1}M < 1/2$.

\section{A first order gravity and thermodynamics}

The first order formalism of gravity can be very useful to analyze thermodynamics. This formalism is
reviewed in \cite{Zanelli:2002qm}. In first order gravity the fields are a basis for the cotangent
space, called the vielbein $e^{a}$, and a connection, $\omega^{ab}$, for the local Lorentz group of
the tangent space. Either $e^{a}$ and $\omega^{ab}$ are understood as differential forms. The
curvature of the Lorentz connection reads,
\[
R^{ab}= d\omega^{ab} + \omega^{a}_{\hspace{1ex} c}\omega^{cb} = \frac{1}{2}
R^{ab}_{\hspace{2ex}
cd} e^{c} \wedge e^{d},
\]
where $R^{ab}_{\hspace{2ex} cd}$ is the Riemann tensor. From now on the $\wedge$
product will be understood implicitly.

\subsection{Einstein Hilbert action with $\Lambda>0$}

The Einstein Hilbert action in first order formalism reads
\begin{equation}\label{EHactionPositive}
\mathbf{I}_{EH} = \kappa_{1} \int_{\mathcal{M}}( R^{ab}  - l^{-2} e^{a} e^{b})
e^{c_{3}}\ldots
e^{c_{d}}\varepsilon_{abc_{3}\ldots c_{d}},
\end{equation}
where the cosmological constant has been written in terms of the dS radius as
$\Lambda =
(d-1)/((d-2) l^2)$. $\kappa_{1}$ is given by \cite{Crisostomo:2000bb}
\[
\kappa_{1}= \frac{1}{2 (d-2)! G_{1} \Omega_{d-2}}.
\]

The variation of Eq.(\ref{EHactionPositive}) yields the equations of motion,
\begin{equation}\label{EinsteinEquationsPositive}
\left(R^{ab} - \frac{d}{(d-2) l^2} e^{a} e^{b}\right)e^{c_{4}}\ldots
e^{c_{d}}\varepsilon_{cabc_{4}\ldots c_{d}}=0,
\end{equation}
and the equations $T^{a}= de^a + \omega^{a}_{\hspace{1ex} b} e^{b}=0$ which
define a torsion free connection. When the torsion free condition is replaced in
Eqs.(\ref{EinsteinEquationsPositive}) they becomes the standard Einstein equations with a positive
cosmological constant.

As usually the variation of the action also yields a boundary term. This term is
given by
\begin{equation}\label{BTerm}
 \Theta(e^{a},\delta \omega^{ab}) = \left( \delta \omega^{ab} e^{c_{3}}\ldots
e^{c_{d}}\right)\varepsilon_{abc_{3}\ldots
c_{d}},
\end{equation}
and represents the first step to fix the boundary conditions.

For a null or negative cosmological constant $\Theta$ usually diverges, however in this case,
since the space is bounded by the cosmological horizon, is finite. Furthermore, the action itself
Eq. (\ref{EHactionPositive})  is also finite. In principle this makes unnecessary to introduce any
kind of re-normalization process. Because of that the condition
\[ \delta \omega^{ab}|_{\partial
\Sigma}=0
\]
is a sound boundary condition in this case, and it will be chosen in this work.

Returning to the discussion, it is direct to prove that fixing the spin connection at the horizons
determines the temperature of those horizons \cite{Aros:2001gz}. First one must recall that for
stationary black hole the horizon is the surface where the so called horizon generator, the
timelike Killing vector $\xi=\xi^{\mu}\partial_{\mu}$, becomes a light like vector. In
Schwarzschild-dS $\xi=\partial_{t}$. The key relation to obtain the temperature is that at any
horizon $\xi$ satisfies \cite{Aros:2001gz}
\begin{equation}\label{Temperature}
(I_{\xi} \omega^{a}_{\hspace{1ex} b}) \xi^{b}|_{\mathbb{R}\times
\partial\Sigma_{H}} = \kappa \xi^{b},
\end{equation}
where $\kappa$ is the surface gravity at that horizon \footnote{The relation (\ref{Temperature})
is the first order version of the relation
\[
\xi^{\mu}\nabla_{\mu}(\xi^{\nu})|_{\mathbb{R}\times
\partial\Sigma_{H}} = \kappa \xi^{\nu}
\]
obtained in \cite{Wald:1993nt}.}. The temperature is given by $T=\kappa/4\pi$
\cite{hawking}.

For the solution above, Eq.(\ref{SchwdeSitter}), the temperature, as defined by Eq.
(\ref{Temperature}), adopts the form
\begin{equation}\label{TemperatureComp}
 T(r_{H}) = \frac{1}{4\pi}\left. \frac{d f(r)^2}{dr}\right|_{\mathcal{H}}=
\frac{1}{4\pi}\left(\frac{d-3}{r_{H}} -
 \frac{(d-1)}{l^2} r_{H}\right),
\end{equation}
where $r_{H}$ stands for either $r_{+}$ or $r_{++}$ in this case.

\section{Charges and conservation}

Following \cite{Wald:1993nt,Aros:2001gz} in this section the thermodynamics is obtained in terms
of the variation of the Noether charge of the solution. Given a Lagrangian
$\mathbf{L}(\phi,d\phi)$ a symmetry is defined as a change in the field configuration,
$\hat{\delta}\phi$ and $\hat{\delta} x^{\mu}= \chi^{\mu}$, which off-shell produces
\[
\hat{\delta} {\mathbf{L}} = \textrm{E.M}\, \hat{\delta} \phi +
d\Theta(\phi,\hat{\delta }\phi) = d
\alpha,
\]
where E.M. stands for the equations of motion and $\Theta$ for the corresponding boundary term.
Using this one could re-deduce the Noether method and proving that, evaluated on the solution, the
current
\begin{equation}\label{currentdensity}
 \hodge{{\bf J}_\chi} = \Theta(\hat{\delta} \phi,\phi) + I_\chi{\bf L}-\alpha,
\end{equation}
satisfies $d(\hodge{{\bf J}_\chi})=0$.

Since the Lagrangian of any theory of gravity must be invariant under diffeomorphisms,
\textit{i.e.} $\hat{\delta} {\mathbf{L}}= - {\cal L}_\chi {\mathbf{L}}$, the conserved current is
obtained by substituting in Eq.(\ref{currentdensity}) $\hat{\delta} \phi = - {\cal L}_\chi \phi$
and $\alpha=0$. For the Einstein-Hilbert action this current is given by,
\[
\hodge{{\bf J}_\chi} = \kappa_{1} d \left(I_{\chi} ( \omega^{ab})
e^{c_{3}}\ldots
e^{c_{d}}\varepsilon_{abc_{3}\ldots c_{d}}\right).
\]

This current can be used to construct as many charges as Killing vectors the space has
\cite{Aros:2001gz}. To analyze the evolution of these charges one can use the approach developed
in \cite{Lee:1990nz} in terms of the so called presymplectic form
\begin{equation}\label{Symplecticmatrix}
  \Xi(\phi,\delta_1\phi,\delta_2\phi) = \int_\Sigma \delta_1
\Theta(\phi,\delta_2\phi)-\delta_2
  \Theta(\phi,\delta_1\phi),
\end{equation}
where $\delta_1$ and $\delta_2$ correspond to functional variations of the
fields. The
presymplectic form defines the structure of the space of configurations,
$\mathcal{F}$.   One must
stress that if either $\delta_{1}$ or $\delta_{2}$ are symmetries then $\Xi$
vanishes
\cite{Lee:1990nz}.

Obviously the space of classical solutions, denoted $\bar{\mathcal{F}}$, is a subspace of
$\mathcal{F}$. Precisely, the evolution of the Noether charges in $\bar{\mathcal{F}}$ defines the
thermodynamics \cite{Iyer:1994ys}. To study their evolution is necessary to introduce a variation
along the parameter of the solutions denoted $\tilde{\delta}$. This yields
\[
\tilde{\delta} \hodge{{\bf J}_\chi} = \tilde{\delta} \Theta(-{\cal L}_\chi \phi, \phi) + I_\chi
d\Theta(\tilde{\delta}\phi,\phi).
\]
This expression can be rewritten, using the relation $I_\chi d= \lie_\chi - dI_\chi$, as
\begin{equation}\label{ExtensionofCharge}
\left( \tilde{\delta} \Theta(-{\cal L}_\chi \phi,\phi)+ \lie_\chi
\Theta(\tilde{\delta}\phi,\phi)\right) = \tilde{\delta} \hodge{{\bf J}_\chi} +
dI_\chi
\Theta(\tilde{\delta}\phi,\phi).
\end{equation}
One can notice that the left hand side, upon integration, is the presymplectic form
$\Xi(\phi,\tilde{\delta}\phi,-\lie_\chi \phi)$, therefore
\begin{equation}\label{VariationofTheCharge}
  \Xi(\phi,\tilde{\delta}\phi,-\lie_\chi \phi)= \int_\Sigma \tilde{\delta}
\hodge{{\bf J}_\chi} + dI_\chi \Theta(\tilde{\delta}\phi,\phi),
\end{equation}
which must vanish since $\delta_{\chi}=-{\cal L}_\chi$ is a symmetry.

The thermodynamical relations arise from this expression evaluated on $\xi$. The right
side of Eq.(\ref{VariationofTheCharge}) for $\xi$ is given by,
\[
\Xi(\phi,\tilde{\delta}\phi,-\lie_\chi \phi)= 0 =\kappa_{1} \tilde{\delta}\int_{\partial
\Sigma}I_{\xi} ( \omega^{ab}) e^{c_{3}}\ldots e^{c_{d}}\varepsilon_{abc_{3}\ldots c_{d}},
\]
where the surface is $\partial\Sigma=\partial \Sigma_{+} \oplus \partial \Sigma_{++}$.
Therefore, the expression turns out to be
\begin{widetext}
\[
\kappa_{1} \tilde{\delta}\int_{\partial \Sigma_{++}}I_{\xi} ( \omega^{ab})
e^{c_{3}}\ldots
e^{c_{d}}\varepsilon_{abc_{3}\ldots c_{d}} =  \kappa_{1} \tilde{\delta}
\int_{\partial \Sigma_{+}}
I_{\xi} ( \omega^{ab}) e^{c_{3}}\ldots e^{c_{d}}\varepsilon_{abc_{3}\ldots
c_{d}},
\]
\end{widetext}
which, since $\omega^{ab}$ is fixed by the boundary conditions, yields the relation between the
fluxes of heat at both horizons,
\begin{equation}\label{TdS}
     T_{++} \delta S_{++} = T_{+} \delta S_{+}.
\end{equation}
Here it has been identified $ \delta S = 4\pi\kappa_{1} \delta \left( A \right)$, where $A$
stands for the area of any of the horizons. The usual $S=A/4$ is obtained using the standard units
(see the discussion in \cite{Crisostomo:2000bb}).

One can notice, by using Eq.(\ref{TemperatureComp}), that $T_{++}< 0$. This only
due to the orientation of the radial normal vectors, which were defined parallel for both horizon
and not inward. Therefore, in Eq.(\ref{TdS}) is accounted with a positive sign the emissions from
the black horizon but
accounted with a negative sign the emissions of the cosmological horizon. Equivalently, one can
recall the Euclidean language and notice that the time coordinate, which here is an
\textit{angle}, has been taken globally anticlockwise, but it should be taken clockwise at the
cosmological horizon to preserve an inward orientation \cite{Gomberoff:2003ea}. Therefore, in
thermodynamical terms the correct temperature of the cosmological horizon is given by
$T_{++}^{c}=-T_{++}$.

Although Eq.(\ref{TdS}) shows the fluxes of heat, this does not give information about the
evolution yet. To address that one must compute the heat capacities of each horizon. Generically
the heat capacity is given by
\[
C = \frac{\partial E}{\partial T} = \left(  \frac{\partial r_{H}}{\partial M}
\right)^{-1} \left(
\frac{\partial T}{\partial r_{H}} \right)^{-1}.
\]
However, it is direct to notice that
\[
 \frac{\partial T}{\partial r_{H}} = - \frac{d-3}{r_{H}^{2}} - \frac{d-1}{l^2} <
0,
\]
and therefore the sign of $C$ actually depends only on $\partial r_{H}/\partial
M$. In
\cite{Gomberoff:2003ea} was shown that
\[
 \frac{\partial r_{++}}{\partial M} < 0 \textrm{  and   } \frac{\partial
r_{++}}{\partial M} > 0,
\]
which proves that the cosmological horizon has positive heat capacity.

One can confirm this result by rewriting the heat capacity in terms of the
radii. This is given by
\[
C(r_{H}) = 2\pi r_{H}^{d-2}
\left(\frac{r_{H}^2-r^2_{0}}{r_{H}^2+r^2_{0}}\right),
\]
where $r_{H}$ stands for either $r_{+}$ or $r_{++}$.

Fortunately, the fact that the heat capacity of the cosmological horizon be positive permits to
foresee the evolution of the space in absence of any external source. Taking the correct signs for
the temperatures one can notice that, for a given value of $M$,  $T_{+} > T_{++}^{c}$. Therefore,
during their interaction due to its positive heat capacity the cosmological horizon would increase
its temperature, and so its radius. Conversely, the black hole horizon would become even hotter
because of its negative heat capacity and shrink. In this way, there should be a net flux of
energy from the black hole horizon into cosmological horizon. In principle this process should not
stop until the complete evaporation of the black hole.

Although the complete description of the final stage of the evaporation, when the temperature of
the black hole diverges, probably would be only obtained when a theory of quantum gravity truly
exist, still one can expect that the final outcome of this process be the de Sitter space.

\section{Decaying process}
The second law of thermodynamics allows to confirm the evolution of the Schwarzschild-dS solution.
Since each horizon has its own entropy \cite{Astefanesei:2003gw}, in this case the second law of
thermodynamics (\ref{SchwdeSitter}) implies the relation between the areas of the horizons
\begin{equation}\label{SLApllied}
    \delta A_{+} + \delta A_{++} \geq 0,
\end{equation}
which in terms of the radii can be rewritten as
\begin{equation}\label{variationOfAreas}
 r_{+}^{d-3} \delta r_{+} +  r_{++}^{d-3} \delta r_{++} \geq 0.
\end{equation}

However, it is straightforward to prove the variation satisfy, since $r_{+}$ and $r_{++}$ are not
independent, that
\begin{equation}\label{VariationofTheRadius}
 \delta r_{+}= -  F(r_{+},r_{++})\delta r_{++},
\end{equation}
with $F(r_{+},r_{++})>0$. The exact expression of $F(r_{+},r_{++})$ can be obtained from
differentiating the relation between $r_{+}$ and $r_{++}$ in the corresponding dimension (See
Eq.(\ref{dDimensionalrelation})). For $d=4$ the relation reads
\[
 \delta r_{+} = -\left(\frac{2r_{+}+ r_{++}}{2r_{++}+
r_{+}}\right)  \delta r_{++},
\]
which allows to rewrite Eq.(\ref{variationOfAreas}) as
\begin{equation}\label{finalVariationofRadii}
\frac{(r_{++}^{2} -  r_{+}^{2})}{(2r_{+}+ r_{++})} \delta r_{++} \geq 0
\Leftrightarrow
\frac{(r_{+}^{2} -  r_{++}^{2})}{(2r_{+}+ r_{++})} \delta r_{+} \geq 0.
\end{equation}
This result determines that the radius of the cosmological horizon must expand, or equivalently
the radius of the black hole must decrease, in order to the second law of thermodynamics be
satisfied. Analogously, for $d=5$ the eq.(\ref{VariationofTheRadius}) reads
\[
 \delta r_{+} = -\left(\frac{r_{++}}{r_{+}}\right)
 \delta r_{++}
\]
and thus Eq.(\ref{variationOfAreas}) in this case reads $(r_{++}^2 -  r_{+}^2) \delta r_{++} \geq
0$, which also implies that the radius of cosmological horizon increases.

After a straightforward, but cumbersome, computation one can prove that in higher dimensions,
using relation (\ref{dDimensionalrelation}), the same result stands, and the radius of the
cosmological horizon must expand due to the second law of thermodynamics.

This result is extremely powerful and general since is based only on the laws of thermodynamics.

\section{Other theories of gravity}

In higher dimensions there are several possible proper theories of gravity \cite{Zanelli:2002qm},
and in principle one could extend some of thermodynamical definitions above to them. For instance
the thermodynamics for Gauss-Bonnet theory is discussed in \cite{Cai:2003gr}. Both Einstein and
Gauss Bonnet theories belong to a lager family of theories called Lovelock gravities, whose
thermodynamics has been also discussed in several articles.

To narrow the possible theories one can requests to have a single positive cosmological constant,
and so avoiding to deal with several different ground states. Within the so called Lovelock
gravities is possible to define a family of theories satisfying that. The Lovelock Lagrangian
 is given by \cite{Lovelock:1971yv}
\begin{equation}\label{Lovelock}
    \mathbf{L} = \kappa \sum_{p=0}^{[(d-1)/2]} \alpha_{p} (R)^{p} (e)^{d-2p}
\varepsilon
\end{equation}
where $(R)^{p} = R^{a_{1} a_{2}}\ldots R^{a_{2p-1} a_{2p}}$,
$(e)^{d-2p}=e^{a_{2p+1}}\ldots
e^{a_{d}}$ and $\varepsilon=\varepsilon_{a_{1}\ldots a_{d}}$. $[(d-1)/2]$ stands
for the integer
part of $(d-1)/2$ and
\[
\kappa_{k}= \frac{1}{2 (d-2)! G_{k} \Omega_{d-2}}.
\]

By a direct translation of \cite{Crisostomo:2000bb} one can determine the relation between the
coefficients that yields a single cosmological constant. Provided
\[
\alpha_{p} = \kappa \frac{(-l^{2})^{p-k}}{d-2p}\left(\begin{array}{c}
k \\
p \\
\end{array}\right)
\]
for $p\leq k$ and $\alpha_{p}=0$ for $p>k$ the action (\ref{Lovelock}) yields
$T^{a}=0$ and the
equations of motion
\[
\left(R-\frac{e^2}{l^2}\right)^{k} (e)^{d-2k-1} \varepsilon = 0.
\]
This confirms the presence of a single positive cosmological constant. These theories of gravity
are usually called $k$-gravities.

The theories above have a solution of the form of Eq.(\ref{SchwdeSitter}) with
\begin{equation}\label{fktheories}
f(r)^2 = 1 -\frac{r^2}{l^2} -
\left(\frac{2MG_{k}}{r^{d-2k-1}}\right)^{\frac{1}{k}}.
\end{equation}

As previously, to avoid naked singularities and to ensure reality $M>0$. Also, one can notice that
when  $d-2k-1=0$ solution presents a naked singularity and thus it will not be considered either.
For $d-2k-1 > 0$ the function $f(r)^2$ may have none, one or two positive solution. As previously
only the case with two horizons, called respectively $r_{+}$ and $r_{++}$, will be considered. In
this case the ranges of those radii are given by $0 \leq  r_{+} < r_{0}$ and $r_{0} < r_{++} \leq
l$ with
\[
r_{0}= l\sqrt{\frac{d-2k-1}{d-1}} .
\]
In addition $0\leq M < M_{max}$ where
\[
  G_{k}M_{max} = \frac{1}{2}r_{0}^{d-2k-1} \left[1 -
\frac{r_{0}^2}{l^2}\right]^{k}
\]

The definition of the temperature, since is purely geometrical, can be obtained from
Eq.(\ref{Temperature}), which in this case reads,
\[
T = \frac{1}{4\pi l^2 k r_{H}}( r_{0}^{2}-r^{2}_{H}),
\]
where $r_{H}$ stands for either $r_{+}$ or $r_{++}$.

The heat capacity can also be computed in this case and it is given, in terms of
the radii, by
\begin{equation}\label{LovelockHeatCapacity}
C_{k}(r_{H}) =  2\pi k r_{H}^{d-2k} \left[1 - \frac{r_{H}^2}{l^2}\right]^{k-1}
 \frac{r_{H}^2-r^2_{0}}{r_{H}^2+r^2_{0}}.
\end{equation}
It is direct from this expression (\ref{LovelockHeatCapacity}) to notice that $C_{k}(r_{++}) >0 $
and $C_{k}(r_{+})<0$. Using the same arguments as for the Einstein theory, one can argue that the
evolution of these black holes should be their complete evaporation.

The analysis using the presymplectic form is also valid for these theories. In this case this also
yields the relation between the differential of heat at both horizons,
\[
    T_{++} \delta S_{++} = T_{+} \delta S_{+},
\]
where the entropy is given by \cite{Aros:2001gz},
\begin{eqnarray*}
S &=& \beta \int_{\partial\Sigma_{H}} I_{\chi} w^{ab} \tau_{ab} \\
   &=& \kappa l^{d-2k}\sum_{p=1}^{k}
\frac{p(-1)^{p-k}}{d-2p}\left(\begin{array}{c}
k \\
p \\
\end{array}\right)\left(\frac{r_{H}}{l}\right)^{d-2p}.
\end{eqnarray*}
Even though there are some negative signs in this expression one can check that this entropy is an
increasing function of the radius.

Unfortunately, in this case the $1/k$ power in $f(r)^2$ rules out the existence of an analytic
relation between the variations of $r_{+}$ and $r_{++}$. One can obtained it, however, by
numerical methods ( up to eleven dimensions and for $k=2\ldots 5$ ). Moreover, after some long
numerical computations one can prove that the second law of thermodynamics also in this case
determines that the respective radii of black hole horizons must decrease.

\section{Conclusions and prospects}
In this article was argued that the quasi statical evolution of the Schwarzschild de Sitter
solution is the complete evaporation of its black hole. The result was obtained from the analysis
of the heat capacities of the horizons, and independently confirmed by using the second law of
thermodynamics. Although the analysis was not made the extension to the Kerr-dS solution seems
natural, and thus one can conjecture that the evolution of any Kerr-dS solutions is also the
evaporation of their black holes. Remarkably the same result stands for any other Lovelock theory
of gravity with a single positive cosmological constant as well.

However, there are some fundamental question to be addressed in the future. In the picture
described in this article the mass of the black hole is radiated beyond the cosmological horizon.
Unfortunately this picture becomes unclear at the transition between the Schwarzschild-dS and dS
spaces. The open question here is what happens with that energy radiated once the black hole
disappears completely. In the de Sitter space beyond the cosmological horizon there is nothing but
the de Sitter space itself, and thus, \textit{roughly speaking}, the energy can not be hidden
there.

\appendix
\section{d dimensional relation} \label{RecursiveRelation}

The relation between $r_{+}$ and $r_{++}$ for the Schwarzschild-dS black hole is
given in
$d$-dimensions by,
\begin{equation}\label{dDimensionalrelation}
    (r_{+}+r_{++})((r_{+}+r_{++}) + a_{2}) - r_{+} r_{++} = l^{2}
\end{equation}
where $a_{2}$ can be obtained recursively from the relation
\[
  a_{d-i} + (r_{+}+r_{++})a_{d-i-1} - r_{+} r_{++} a_{d-i-2} = 0
\]
with $(r_{+}+r_{++}) a_{d} = r_{+} r_{++} a_{d-1}$ and $a_{1} = (r_{+}+r_{++})$.

\section{Kerr-dS}\label{KerrSection}

The discussion has been centered on Schwarzschild-dS solution. This can be considered not general
enough to be good a probe but it indeed has the structures necessary to address the general
problem presented in this article. For instance, the most general four dimensional solution in
vacuum with positive cosmological constant is the Kerr-de Sitter geometry. This, written
Boyer-Lindquist-type coordinates, is given by the vielbein \cite{Carter:1970ea}
\begin{eqnarray}
e^{3}=\frac{\sqrt{\Delta _{\theta }}}{\Xi \rho }\sin \theta
(adt-(r^{2}+a^{2})d\varphi ), & &
e^{2}=\rho \frac{d\theta }{\sqrt{\Delta _{\theta }}}, \nonumber \\
e^{0}=\frac{\sqrt{\Delta _{r}}}{\Xi \rho }(dt-a\sin ^{2}\theta d\varphi ), & &
e^{1}=\rho
\frac{dr}{\sqrt{\Delta _{r}}}\label{KerrTetrad}
\end{eqnarray}
with $\Delta _{r}=(r^{2}+a^{2})\left( 1-\frac{r^{2}}{l^{2}}\right) -2Mr$,
$\Delta _{\theta
}=1+\frac{a^{2}}{l^{2}}\cos ^{2}\theta $, $\Xi =1+\frac{a^{2} }{l^{2}}$ and
$\rho
^{2}=r^{2}+a^{2}\cos ^{2}\theta $.

The horizons in this case are given by the roots of $\Delta_{r}=0$. Moreover, as for the
Schwarzschild-dS solution, the region of interest is defined between the two largest positive
roots, $r_{++}$ and $r_{+}$, which define the cosmological and black hole horizons respectively.
There is another internal horizon in this case \cite{Dehghani:2001kn}, though. It is direct to
prove that those radii are also bounded as $r_{+}<r_{0}$ and $r_{0}< r_{++}< l$ with
\[
r_{0} = \frac{1}{6}\sqrt{6\left(l^2-a^2+ \sqrt{a^4 - 14 a^2 l^2 + l^4}\right)}.
\]

In higher dimensions the Kerr-dS solution has the same generic form of Eq.(\ref{KerrTetrad})
\cite{Dehghani:2001kn} with
\[
\Delta _{r}=(r^{2}+\sum_{l} a^{2}_{l})\left( 1-\frac{r^{2}}{l^{2}}\right)
-2Mr^{5-d},
\]
where $a_{l}$ are the coefficients related with the angular momenta in higher dimensions. This
function also defines two horizons.

These analogies with the Schwarzschild-dS solution confirm that this solution is enough general to
address the general problem properly. Of course the transmission of heat in Eq.(\ref{TdS}) should
be modified by the presence of angular momenta or electric charge, nonetheless the second law of
thermodynamics, which depends only on the radii, should be reducible to the form
Eq.(\ref{VariationofTheRadius}).

{\bf Acknowledgments}

I would like to thank D. Astefanesei for calling my attention to his work. I also would like to
thank Abdus Salam International Centre for Theoretical Physics (ICTP). This work was partially
funded by grants FONDECYT 1040202 and DI 06-04. (UNAB)

%\bibliographystyle{JHEP}
%\bibliography{myXbib}
\providecommand{\href}[2]{#2}\begingroup\raggedright\endgroup

\end{document}